\documentclass[12pt]{iopart}
  \expandafter\let\csname equation*\endcsname\relax
  \expandafter\let\csname endequation*\endcsname\relax
\usepackage{amsmath, amssymb}
 \usepackage{graphicx}
 \usepackage{amsmath}
 \usepackage{amssymb}
  \usepackage{enumerate}
\usepackage{cite}
\usepackage{hyperref}

\newcommand{\be}{\begin{eqnarray}}
\newcommand{\ee}{\end{eqnarray}}

\begin{document}

 \title[Relativistic Particle Motion and Quantum Optics in Weak Gravity]{Relativistic Particle Motion and Quantum Optics \\ in a Weak Gravitational Field}
\author{C. Anastopoulos$^1$ and Bei-Lok Hu$^2$}

\address{$^1$Department of Physics, University of Patras, 26500 Patras, Greece.}

\address{$^2$Maryland Center for Fundamental Physics and Joint Quantum Institute,\\ University of
Maryland, College Park, Maryland 20742-4111 U.S.A.}

\ead{anastop@upatras.gr,blhu@umd.edu}

\begin{abstract}
The possibility of long-baseline quantum experiments in space makes it necessary to better understand the time evolution of relativistic quantum particles in a weakly varying gravitational field.  We explain why conventional treatments by traditional quantum optics and atomic physics  based on quantum mechanics, may become inadequate when faced with issues related to locality, simultaneity, signaling, causality, etc.  Quantum field theory which respects the principles of special relativity  is needed.   Adding to this the effects of gravitation, we are led  to Quantum Field Theory in Curved Spacetime (QFTCST) as the  only  consistent theory for describing quantum field theoretic and general relativistic effects.  This well-established theory should serve as the canonical reference  theory to a large class of proposed space experiments testing the foundations of gravitation and quantum theories, such as the equivalence principle  for quantum systems, and the basic notions of quantum information theory in relativistic settings.  This is the first in a series of papers treating near-term quantum optics and matter waves experiments in space from the perspective of QFTCST.  Here, we analyze the quantum motion of photons and of scalar massive particles using QFTCST with application to interferometer experiments. Our main result is that, for photons, the weak gravitational field is to leading order completely equivalent to an inhomogeneous dielectric, thus allowing for a description of quantum optics experiments in curved space using familiar notions from the theory of optical media. We also discuss interference experiments that probe first-order quantum coherence, the importance of a covariant photo-detection or particle detection theory, and the relevance of time of arrival measurements. For massive particles with internal structure, we identify a novel gravity-induced phase shift that originates from the different gravitational masses attributed to the excited internal states. This phase shift can in principle be measured in space experiments.
\end{abstract}
 \maketitle

\section{Introduction}

This paper treats relativistic quantum particle motion in a weak gravitational field, a topic which bears on the broad subject of quantum and atomic optics  (Q/AOpt) under gravity, on Earth or in its nearby space extending to the whole solar system.  Our work is motivated by two developing projects, one is theoretical:  we need this to extend our recently obtained results on the equivalence principle for quantum systems (EPQS) \cite{EPQS} to relativistic systems, for massive particles with internal structure (fast moving atoms) and for photons; the other is experimental: we want to develop enough basic theory to interface with the NASA-JPL Deep Space Quantum Link (DSQL) \cite{DSQL, Makan} initiative in carrying out long baseline quantum experiments between the  Earth and the Moon. See also the proposal  of Ref. \cite{Terno}, and the earlier proposal of Ref. \cite{Rideout} that covers a broader range of theoretical issues.

Both aspects have a long history of development spanning more than half a century. We shall only mention some representative work on some nodal themes.    On the theory-motivated side,  relativistic quantum mechanics  \cite{Ohlsson} lacks full consistency and generality. The ultimate need will eventually be felt to use quantum field theory (QFT) in curved spacetime (CST) \cite{BirDav, Fulling, Wald} in order to formulate quantum optics and quantum information theories in a gravitational field setting, as explained below.  This paper takes a small step in that direction.  See also Refs. \cite{RelQOp1, RelQOp2, RelQOp3} with similar concerns.

On the experiment-motivated side, there is ample work   in quantum optics, starting with the tests of classical general relativity \cite{Will} to the famous Collela-Overhauser-Werner (COW) experiments \cite{COW}, from the use of laser interferometry to atom interferometry \cite{Borde, LamBor, MarAud, Roura} to neutron interferometry \cite{neutron, Ana, Abele, Ryder, Green}. We shall only consider scalar and electromagnetic fields  in this work, describing scalar particles and photon optics, but we want to mention that future quantum information experiments using relativistic particles with spin \cite{Mash, HehlNi, KhrPom, Alsing}, including frame dragging effects,  may provide even more complex probes into the interplay of quantum matter and the gravitational field,  both native, as in the Earth's environment or distant, as from incoming gravitational waves carrying information about  black hole and neutron star activities afar.

\subsection{Upgrading the theoretical basis of quantum optics under gravity to QFTCST}

We begin by elaborating on this preamble message mentioned above,  suggesting a paradigmatic change, that  for the analysis of quantum effects in space experiments, both in quantum optical and matter-wave set-ups,  quantum mechanics is not enough,  QFTCST is needed.  Let us focus on the field aspect which is inadvertently yet generally played down in quantum optics.

\paragraph{Quantum Correlations at Spatial Separations.}  Experiments  involving correlations of detection events located at different spatial separations need QFT. For example, it is possible to measure the joint probability of particle detection $p(X_1, X_2)$ at spacetime points $X_1$ and $X_2$, a quantity closely related to the second order coherence of a field.   In quantum optics, the theoretical determination of such correlations requires the development of a photodetection theory. The  best known such theory is Glauber's \cite{Glauber}, which expresses quantities such as $p(X_1, X_2)$ in terms of field correlation functions. Hence, physical predictions for such experiments require  an explicit QFT description that also incorporates the effects of field propagation in a gravitational background, invoking CST.

Current Bell tests  appear   to not need QFT, but a QFT treatment may be necessary in future tests with moving detectors \cite{DSQL}, and when keeping track of other particle observables (e.g., time of arrival) beyond the ones of the entangled qubits.

\paragraph{Causality and Locality Issues.}
The study of correlations at spatially separated points raises forcefully the issue of {\it causality}. Glauber's theory has problems with this issue, as it employs a version of the Rotating Wave Approximation that typically misrepresents retarded propagation \cite{RWA}. An upgrade of existing photo-detection theory is needed in order to address quantum optical experiments in space.  Analogous developments are necessary for experiments with massive particles.

This elicits foundational issues of QFT on {\it locality and transmission of information} that so far have largely avoided direct testing. There is a long-standing tension between the notion of a localized system (a  particle detector, or an emitting atom) and causal propagation of signals that so far has not been resolved. Well-established theorems assert that even unsharp localization of a system in a spatial region leads to faster than light signals \cite{Schl71, Heg98, Mal96}.

\paragraph{Fermi's two-atom problem.}
Perhaps better known is Fermi's two-atom problem  \cite{Fermi}. Fermi assumed that at time $t = 0$,  atom A is in an excited state and atom B in the ground state, the two atoms being separated by distance $r$. He asked when B will notice A and move from its ground state. In accordance with Einstein locality, he found that this happens only at times $t  \geq r$. It took about thirty years for Shirokov to point out that Fermi's result is an artifact of an approximation \cite{Shiro}.

Later studies reached   conclusions that depended on the approximations used \cite{Fermiproblem}. Eventually, Hegerfeldt showed that non-causality is generic \cite{Heg,Heg2}, as it depends only on the assumption of energy positivity and on the localization of the atoms in disjoint spacetime regions---see, also the critique in \cite{Buch} and a recent exactly solvable model \cite{RelQOp2}.

\subsection{Space experiments on foundational issues of QFT, QI and Gravitation}

The two-atom problem is an example of a genuinely foundational and physically meaningful problem of QFT, that pertains to the  meaning of locality in relation to quantum measurements, which can be addressed by space experiments in the near future.

\paragraph{Quantum Information (QI).}
Since  quantum experiments in space are due to explore quantum informational aspects like entanglement in a relativistic setting, the incorporation of quantum information concepts in QFT is a  theoretical necessity.  So far, quantum information theory has been largely developed in the context of non-relativistic quantum mechanics, a small corner of full QFT.  It is ostensibly inadequate when basic relativistic effects – both special and general -- such as causality and covariance, need be accounted for.

\paragraph{Spacetime Structure and Gravitation.}
QFT is embedded with axioms about the effects of spacetime structure on the properties of quantum systems, especially regarding the causal propagation of signals.  In contrast, current QI theories do not incorporate the latter axioms. Their notion of causality, based on the sequence of successive operations on a quantum system---see, for example, \cite{causalQI}---lacks a direct spacetime representation. As a result, current theory does not make crucial relativistic distinctions, for example, between timelike and spacelike correlations, it does not describe real-time signal propagation, and it ignores relativistic constraints on permissible measurements---for some exceptions, see Refs. \cite{Relcaus}. To overcome such limitations and to eradicate these pathologies a  relativistic QI theory need be formulated from first principles and to be based on quantum field theory.

Furthermore, experiments that study the gravitational interaction in multi-partite quantum systems require a QFT treatment of interactions for consistency.  A non-QFT description may severely misrepresent the theoretical modeling of the system or the physical interpretation of the results. This point is crucial, for example, in the search of EPQS  \cite{EPQS, OnVi, ZyBr},  in experiments with gravitational cat states \cite{AnHu15, Bose17, Vedral17, AnHucr,AnHu20},  in the search of indefinite causal order due to gravity \cite{ZCPB}, or manifestations of the problem of time in the weak gravity limit \cite{ALS21}.

\subsection{Our Intents  and our Findings}

\paragraph{Our Intents.}

To address the above concerns, there is a clear need to begin building  bridges between the QFTCST community and the quantum and atom optics (Q/AOpt) community, both  established  since the 60's and widely applied:  the former mostly to strong field settings such as for black hole and early universe physics, while the latter mostly to low energy tabletop experiments, extending to space  in the recent decades.  There is also some conventional bias in that QFTCST remains theory-laden with a scarcity in experiments, while advances in Q/AOpt  are mainly experimentation and instrumentation driven. A welcoming development in the last two decades is that three emergent fields, namely, analogy gravity \cite{AnalogG}, gravitational quantum physics \cite{GQP}, and relativistic quantum information \cite{RQI}, are tapping into these two sources --QFTCST and Q/AOpt -- in productive ways.  We hope our work can add some bricks to the construction of necessary bridges between them.

\paragraph{This work.}

We formulate QFTCST in a weak static gravitational field emphasizing the aspects that are relevant to the planned deep space experiments, while translating QFTCST concepts and methods to the quantum  optics / matter-waves language. Since QFTCST is the only consistent theory currently available for describing particle propagation in gravity, every physical prediction must be eventually phrased in this language. This is particularly important for precision measurements and for tests of fundamental principles (locality, causality, equivalence principle, and so on).

Our results are the following.

\medskip

\noindent  1.  We show  that the Newtonian gravitational field $\Phi$ acts---to leading order---like an inhomogeneous dielectric on Minkowski spacetime with a refractive index $n = 1 +  2|\Phi|$. This remarkable fact provides a mapping from the description in terms of \textit{curvature} to one in terms of \textit{optics in a medium}, that of a quantum field in presence of dielectric. In the weak field regime the calculations, without relying on differential geometry,   are thus made simpler and the physical picture more transparent.

The mathematical analogy between curved spacetime and a dielectric medium has been long known in classical general relativity \cite{Skrotski, Plebanski, defelice}.  However, to the best of our knowledge, {\it the simplicity of the gravity-induced phase shift} at weak gravitational fields, its independence from wavelength, and its persistence in the quantum regime, has not been pointed out before. For experiments within the Earth-Moon-Sun system, where the gravitational potential has a very complex inhomogeneous behavior, the conceptual and technical simplicity afforded by the relation we obtained, is substantial.

In special cases, our results coincide with previously derived results for the  phase change due to the propagation in a gravitational field, see, for example, \cite{Terno, terno2}. However, our derivation  is fully quantum and only involves a perturbation in the strength of the potential, it does not require a semi-classical, or a fully classical geometric-optics approximation.

\medskip

    \noindent 2. We then undertake {\it the analysis of interference experiments}. Our methods are rather standard, but the context is relatively novel. Our analysis primarily serves as a methodological template for more complex calculations in the future. We place strong emphasis on the fact that {\it physical predictions strongly depend on the photo-detection theory}. Here, we only use the standard Glauber's method, pointing out that an alternative approach, the Quantum Temporal Probabilities (QTP) method \cite{AnSav12, AnSav19}, leads here to the same results. The identification of differences requires the treatment of higher order coherences, which we leave for future work.

\medskip

\noindent  3.  Using the same method, we also identify {\it the gravity induced rotation of polarization} \cite{polar}. In the context of the interferometry experiments considered here, this rotation is manifested as a loss of visibility.

 \medskip

\noindent 4.  We  show that the change in optical length due to gravity can  be measured in time-of-arrival measurements. We point out that it is impossible to distinguish the gravitational contribution to the interference phase in experiments using only photons, because photons `see' the same distances in all experiments. The length of the interferometric arms have to be determined through other methods, i.e., particles that do not move on the lightcone.

\medskip

 \noindent 5.  As we are interested in {\it matter wave interferometry}, in addition to optical interferometry, we also treat the case of  {\it massive spinless particles}. The perturbative calculation of the phase shift is less reliable here, and it applies only for sufficiently large momenta. For slow moving particles, the effect of gravity is significantly more complex than a phase shift. A semiclassical approximation (e.g., WKB) may provide a practical method for computing the field modes, but any discussion of the equivalence principle of quantum systems must be made on the basis of exact solutions.

\medskip

 \noindent 6. For massive particles with internal degrees of freedom, we identify a phase shift that originates from the change in the gravitational mass of the particle when an internal degree of freedom is excited. This phase shift generalizes the one of Ref. \cite{EPQS} that was derived for non-relativistic particles in a homogeneous gravitational field. This phase shift is too small to be measured in Earth-based experiments; however, it can plausibly be measured in a space setup with long baselines.

\bigskip

This paper is organized as follows: In Sec. 2, we consider a quantum scalar field in weak gravity, where we identify the field modes perturbatively. Then, we treat the case of massive particles through the WKB approximation, and identify the gravity induced phase shift for composite particles.
 In Sec. 3, we repeat this analysis for the quantum EM field. In Sec. 4, we discuss photodetection theories and their relevance to physical predictions. In Sec. 5, we discuss different interferometric schemes and time of arrival measurements. In Sec. 6, we place our results in the broader context of the necessity of QFTCST for relativistic quantum optics.

\section{Quantum scalar field in  weak gravity}

\subsection{Constructing QFTs for free fields}
The construction of QFTCS is obstructed by the fact that the Poincar\'e group, so crucial to QFT in Minkowski spacetime, is absent in generic spacetimes. However, for free fields there is a well defined quantization procedure that proceeds through the following steps \cite{Wald}.
\begin{enumerate}
\item Find all solutions to the classical field equations for the fields $\phi^A(X)$, where $A$ is a collective index.
\item Identify a complex-valued set of modes $f_n^A(X)$ as having `positive frequency'; $n$ is an index that labels the modes.  Any solution to the field equations can be written as $\sum_n [a_nf_n^A(X)+a^*_n\bar{f}^A_n(X)]$ for some complex numbers (amplitudes) $a_n$.
\item An appropriate inner product $(\cdot , \cdot)$is introduced so that     $(f_n, f_m) = \delta_{nm}$.
    \item Promote the amplitudes $a_n$ to annihilation operators $\hat{a}_n$ and their conjugates to creation operators $\hat{a}^{\dagger}_n$ on a Fock space that may be either bosonic or fermionic. The vacuum state is specified by $\hat{a}_n|0\rangle = 0$ for all $n$.
\item The Heisenberg-picture quantum field $\hat{\phi}^A(X)$ is expressed as
\be
\hat{\phi}^A(X) = \sum_n\left[ \hat{a}_nf_n^A(X)+ \hat{a}^{\dagger}_n\bar{f}^A_n(X) \right].
\ee
\end{enumerate}

The non-trivial part in this procedure is the selection of the `positive frequency' modes. In generic spacetimes, no natural choice exists. However, if a spacetime has a preferred global time coordinate $t$,   $f_n^A(X)$ can be selected by the requirement that $i \frac{\partial f^A_n}{\partial t} = \omega_n f^A_n$, where $\omega_n \geq 0$. This is the case for stationary spacetimes, where $\frac{\partial}{\partial t}$ is the timelike Killing field for the spacetime metric.

Even in weak gravity, the spacetime geometry in the near Earth environment is not stationary because of motion in the Earth-Moon system. Let $\omega_0$ stand for  the characteristic frequencies of the latter. For field frequencies $\omega$ such that $\omega_0/\omega << 1$, the time-dependence of the geometry can be treated in the adiabatic approximation. Assuming a stationary spacetime metric as the leading order approximation is well justified.  Furthermore, the phenomena induced by Earth's rotation are small compared to the phenomena that are described by a gravitational potential. This fact justifies the  use of  QFTs in a  static (rather than a stationary) spacetime as a background, to which rotational and time-dependent effects can be added as small perturbations.

\subsection{Klein-Gordon equation on a static spacetime}
A scalar field $\hat{\phi}(X)$ in a spacetime $M$ with Lorentzian metric $g_{\mu \nu}$ satisfies the Klein-Gordon equation
\be
\frac{1}{\sqrt{-g}}\partial_{\mu}(\sqrt{-g}g^{\mu \nu}\partial_{\nu} \hat{\phi}) - m^2 \hat{\phi} = 0 ,
\ee
where $g = \det g_{\mu \nu}$.
We specialize to the case of a static spacetime with metric
\begin{eqnarray}
ds^2 = - N({\pmb x}) dt^2 + h_{ij}dx^i dx^j, \label{metric}
\end{eqnarray}
where $N$ is the lapse function and $h_{ij}$ is a Riemannian three-metric.

We then look for positive frequency solutions to the Klein-Gordon equation, i.e., solutions of the form $f_n ({\pmb r}) e^{- i \omega_n t}$, where $\omega_n > 0$. They satisfy
\begin{eqnarray}
\nabla^2 f_{n} + N^{-1}  \nabla_i N  \nabla^if_n  -(  m^2 - \omega_a^2N^{-2}) f_n  = 0. \label{WE}
\end{eqnarray}
The functions are normalized in accordance with the Klein-Gordon inner product
\begin{eqnarray}
\int d^3x \sqrt{h} N^{-1}(x) f_n(x) f^*_m(x) = \delta_{nm}
\end{eqnarray}

The field operator is $\hat{\phi}(t, {\pmb r}) = \sum_n \left[ \hat{a}_n f_n({\pmb r}) e^{- i \omega_n t} + \hat{a}^{\dagger}_n f^*_n({\pmb r})e^{i \omega_n t}\right]$, in terms of the annihilation operators $\hat{a}^n$ and the creation operators $\hat{a}_n^{\dagger}$. The Wightman function for a general field state $|\Psi\rangle$ is
\begin{eqnarray}
G^{(2)}(t, {\pmb x}; t', {\pmb x}') = 2 \mbox{Re} \left( \sum_{ab} \rho^n_m f_n({\pmb x}')f^*_m({\pmb x})e^{-i (\omega_n t - \omega'_m t')} \right) \nonumber \\+ \sum_n f_n({\pmb x}) f^*_n({\pmb x}')e^{- i \omega_n(t-t')},
\end{eqnarray}
 where $\hat{\rho}^n_m = \langle \Psi|\hat{a}^{\dagger}_m \hat{a}^n|\Psi\rangle$ is the one-particle reduced density matrix associated to $|\Psi\rangle$.

 \subsection{Field modes}
 Next, we consider the case of a weak gravitational field that is a solution to Einstein's field equations in vacuum. This implies that
 \begin{eqnarray}
N = e^{\Phi} \simeq 1 + \Phi, \hspace{0.5cm} h_{ij} = e^{-2 \Phi} \delta_{ij} \simeq (1 -  2\Phi) \delta_{ij}, \label{vacugr}
 \end{eqnarray}
where $\Phi$ is the Newtonian gravitational potential. It satisfies $|\Phi| << 1$ and $\nabla^2 \Phi = 0$ with respect to the flat space Laplacian.

To leading order in the potential $\Phi$, Eq. (\ref{WE}) becomes
\begin{eqnarray}
\nabla^2 f + q^2 f = 0. \label{WE2}
\end{eqnarray}
where we dropped the index $n$ for brevity, and wrote
\be
q^2  = \omega^2 e^{-4\Phi} -  m^2 e^{-2 \Phi} \simeq k^2 -2 (m^2+2k^2)\Phi,
\ee
where $k^2 = \omega^2 - m^2$.

For massless particles $q^2 \simeq  k^2 e^{-4 \Phi} \geq 0$. For massive particles, we note that in the non-relativistic regime $q^2 \simeq k^2 - 2m^2 \Phi$, and it may become negative even while $|\Phi |<< 1$.

\subsection{Perturbative evaluation of the modes}

We write $f = Ae^{iS}$, for $A>0$ and real $S$, to obtain
\begin{eqnarray}
\nabla^2 S +  2 A^{-1} \nabla S \cdot \nabla A = 0 \label{WE3a}\\
(\nabla S)^2 = A^{-1}\nabla^2A +q^2 \label{WE3b}
\end{eqnarray}
Eq. (\ref{WE3a}) can be expressed as $\nabla \cdot(A^2 \nabla S) = 0$, with solution
\begin{eqnarray}
A^2 \nabla S = {\pmb b}, \label{WE3c}
\end{eqnarray}
where ${\pmb b}$ is a divergence-free vector field. Different choices of ${\pmb b}({\pmb x})$ correspond to different mode choices, i.e., different bases in the degeneracy subspaces of Eq. (\ref{WE2}).

Then, Eq. (\ref{WE3b}) becomes
\be
\nabla^2A +q^2A - \frac{{\pmb b}^2}{A^3} = 0. \label{WE3d}
\ee

For $\Phi = 0$, we choose the plane-wave modes. They are  labeled by a momentum vector ${\pmb k}$, such that $A = 1$  and $S = {\pmb k} \cdot {\pmb x}$, with $|{\pmb k}| = k$ . For this choice, ${\pmb j} ({\pmb x}) = {\pmb k}$ is constant.

For $\Phi \neq 0 $, we write $q^2 = k^2 - z$, where $z = (4k^2 + 2 m^2)\Phi$  is the correction due to the potential. We also
write $A = 1 + \alpha$,  $S = {\pmb k}\cdot{\pmb x} + \sigma$,  and ${\pmb j} = {\pmb k} + {\pmb b}$ where $\alpha$, ${\pmb b}$ and $\sigma$ are of order $z$. To leading order in $\Phi$, Eq. (\ref{WE3c}) becomes,
\be
\nabla \sigma = - 2 \alpha {\pmb k} + {\pmb b}, \label{WE4a}
\ee
while Eq. (\ref{WE3d}) yields
 \be
\nabla^2 \alpha + 4 k^2 \alpha = z + 2 {\pmb k} \cdot {\pmb b} \label{WE4c}
\ee

The solution to Eq. (\ref{WE4c}) is $\alpha = G(z + 2 {\pmb k} \cdot {\pmb b})$, where $G = (\nabla^2 +4 k^2)^{-1}$. However,  $z$ is proportional to $\Phi$ and $\nabla^2 \Phi = 0$ in vacuum. We can write $G$ as a power series $\sum_{n=0}^{\infty}c_n (\nabla^{2})^{n}$, so that each term in the expansion vanishes when acting on $z$, except $n = 0$. Hence, $Gz = \frac{z}{4k^2}$, and the solution of Eq. (\ref{WE4c}) is
\be
\alpha = \left(1+ \frac{m^2}{2k^2}\right) \Phi + 2 {\pmb k} \cdot (G{\pmb b}).
\ee

Substituting into Eq. (\ref{WE4a})
\begin{eqnarray}
\nabla \sigma = - 2 (1 +\frac{m^2}{2 k^2}) {\pmb k} \Phi + {\pmb \lambda},  \label{WE5a}
\end{eqnarray}
where
\be
{\pmb \lambda} = {\pmb b} - 4 {\pmb k} [{\pmb k} \cdot (G{\pmb b})]
\ee
These solutions depend on the arbitrary  vector field ${\pmb b}$. Any choice of ${\pmb b}$ consistent with the normalization of the modes is acceptable. For propagation in one dimension ${\pmb b}$ vanishes. Furthermore, if we want the modes to correspond to propagation along the direction specified by ${\pmb k}$, then ${\pmb b}$ must be proportional to ${\pmb k}$. But then the requirement that $\nabla \cdot {\pmb b} = 0$ implies that ${\pmb b}$ must be a constant. Then ${\pmb \lambda} = 0$, and ${\pmb b}$ corresponds to $\alpha$ a constant factor. Hence, for linearly propagating modes, we can take ${\pmb b} = 0$ without loss of generality.

To solve Eq. (\ref{WE5a}), we choose a coordinate system so that ${\pmb k} = (0, 0, k)$, and integrate to obtain
\begin{eqnarray}
\sigma(x, y, z)=   \sigma(x,y, z_0)  - 2 k (1 +\frac{m^2}{2 k^2}) \int_{z_0}^z dz' \Phi(x, y, z')
\end{eqnarray}
Since $\sigma(x,y, z_0)$ is arbitrary, we can always choose the modes so that it vanishes. Hence,
\begin{eqnarray}
\sigma(x, y, z)=    - 2 k (1 +\frac{m^2}{2 k^2}) \int_{z_0}^z dz' \Phi(x, y, z') \label{phase}
\end{eqnarray}
For brevity, we write Eq. (\ref{phase}) as
\be
\sigma = -2  (1 +\frac{m^2}{2 k^2}) {\pmb k} \cdot \int  \Phi d{\pmb x}.
\ee
Then, the mode solutions are
\begin{eqnarray}
f_{\pmb k}({\pmb x}) =  \frac{1}{(2\pi)^{3/2}} e^{ (1 +\frac{m^2}{2 k^2}) \Phi + i {\pmb k} \cdot ( {\pmb x} - \delta {\pmb x})} \label{modesol}
\end{eqnarray}
where
\be
\delta{\pmb x} = -2  (1 +\frac{m^2}{2 k^2})\int  \Phi d{\pmb x}.
\ee

For $m = 0$, the modes (\ref{modesol})  are orthonormal to leading order in the potential $\Phi$. For $m\neq 0$, the orthonormality condition holds only up to zero-th order. Here, we will focus on the case $m = 0$, Interestingly, a slight modification of the modes (\ref{modesol}) to
\begin{eqnarray}
f_{\pmb k}({\pmb x}) =  \frac{1}{(2\pi)^{3/2}} e^{ (1 -\frac{m^2}{2 k^2}) \Phi + i {\pmb k} \cdot ( {\pmb x} - \delta {\pmb x})} \label{modesol2}
\end{eqnarray}
satisfies orthonormality up to first order in the perturbation. However, Eq. (\ref{modesol}) is perturbatively accurate only at the phase and not at the amplitude for $m = 0$. Either (\ref{modesol}) or (\ref{modesol2}) can be used if we are interested only in the phase shift of matter waves, and for sufficiently large particle momentum so that $\frac{m^2}{2 k^2}\Phi << 1$. Otherwise, a different approximation must be used, like the WKB approximation below.

For $m = 0$, the field describes the scalar analogues of photons. In this case,  the term $\delta {\pmb x}$
 can be interpreted as a change in optical length. This change in optical length is equivalent to that of light traversing a dielectric medium with index of refraction
\be
n = 1 + 2  |\Phi|
\ee
Note that $n >1$ because $\Phi < 0$.

This result is compatible with the following fact \cite{ACL88}. For any static metric (\ref{metric}), the spatial trajectories of massless particles correspond to geodesics of the optical metric $\tilde{h}_{ij} = L^{-2}h_{ij}$. In the present context, $\tilde{h}_{i} = e^{-4 \Phi} \delta_{ij}$. Indeed, the optical length derived by the modes (\ref{modesol}) is the proper length of the optical metric.

For example, consider a  homogeneous gravitational field with potential $\Phi = - g x$. The phase change  for $x$ varying from $x = 0$ to $x = h$
\begin{eqnarray}
\Delta \varphi = 2 \omega \int_0^h dx gx = \omega gh^2.
\end{eqnarray}

\subsection{The WKB approximation for massive particles}
The modes (\ref{modesol}) are derived in the regime $(1+\frac{m^2}{2k^2})|\Phi| << 1$. For $m \neq 0$, this approximation may fail  if $k << m$. In this regime, a semi-classical approximation is more appropriate. We assume that the potential varies at a scale much larger than the wavelength, so that we can drop the term involving $A$ from Eq. (\ref{WE3b}). Then, $|\nabla  S|^2 = q^2$, and we can write $\nabla S = q {\pmb n}$ for some unit vector ${\pmb n}$. Then,
\be
S = {\pmb k} \cdot \int d {\pmb x} \sqrt{1 - 4 (1 + \frac{m^2}{2k^2})\Phi},
\ee
where ${\pmb k} = k {\pmb n}$. We also obtain $A = [1 - 4(1 + \frac{m^2}{2k^2})\Phi]^{-1/4}$.

Hence, the modes are
\be
f_{\pmb k}({\pmb x}) = \frac{D}{(2\pi)^{3/2} [1 - 4 (1 + \frac{m^2}{2k^2})\Phi]^{1/4}} e^{i {\pmb k} \cdot \int d {\pmb x} \sqrt{1 - 4 (1 + \frac{m^2}{2k^2})\Phi}},  \label{wkbmode}
\ee
in terms of a normalization constant $D$.

For $(1+\frac{m^2}{2k^2})|\Phi| << 1$, the modes (\ref{wkbmode}) reduce to (\ref{modesol}). In particular, this is the case for $m = 0$. Note, however, that the modes (\ref{modesol}) do not follow from the WKB approximation, but solely from the requirement of a weak potential $\Phi$.

Eq. (\ref{wkbmode}) applies to the classically allowed region where $q^2 > 0$. An extension to $q^2 = 0$ is possible using the  WKB connection formulas. However, in the present context, the particle de Broglie wavelength is much smaller than the length scale of change in the potential, so that in most cases tunneling is negligible. Then, we are justified in using the approximation where the mode function vanishes in the classically forbidden region. This approximation is sometime called the "geometric-optics approximation" in semi-classical quantum theory, but it does not coincide  with the usual geometric optics approximation of EM theory.

To see that tunneling is negligible, consider the simplest case of particle propagation along the axis that connects the centers of mass of two planetary bodies with mass $M_1$ and $M_2$. Call the distance of the two bodies $R$. Then, the potential along this axis is $V(x) = - \frac{GM_1}{x} - \frac{GM_2}{R-x}$, where $x$ is the distance from the center of mass of the first body. This potential has a maximum at $x = x_m :=  R (1+ \mu)^{-1}$, where $\mu = \sqrt{M_1/M_2}$. We Taylor-expand the potential around $x = x_m$, to obtain an inverse harmonic oscillator potential
\be
\Phi(x) = - \frac{GM_1}{R}(1+\mu)(1+\mu^{-1}) \left[ 1 + (1+\mu)^2 \frac{(x-x_m)^2}{R^2}\right].
\ee
Using the standard WKB formula for tunneling, it
is straightforward to show that the spread of energies $\delta E$ in which the tunneling effect is non-negligible is of the order of $ \sqrt{\frac{GM_1}{R^3}(1+\mu^{-1})(1+\mu)^3}$. For the Earth-Moon system, $\delta E \sim 10^{-20}$eV, hence, the geometric optics approximation of ignoring tunneling is well justified.

\subsection{Massive particles with internal degrees of freedom}

We can also consider the case that the massless particles have internal degrees of freedom that correspond to $N + 1$ different internal states. The associated field $\hat{\phi}_a(X)$   carries an index $a = 0, 1, 2, \ldots, N$. Each component satisfies the Klein-Gordon equation with a different mass
\be
\frac{1}{\sqrt{-g}}\partial_{\mu}(\sqrt{-g}g^{\mu \nu}\partial_{\nu} \hat{\phi}_a) - m_a^2 \hat{\phi}_a = 0.
\ee
We choose the index $a$ so that the masses $m_a$ are non-decreasing functions of $a$. Then, $a = 0$ is the ground state of the internal degrees of freedom, and $m_0$ is the particle's mass in absence of internal excitations. We define $\epsilon_a = m_a - m_0$, and consider the regime where $\epsilon_a << m_0$. We also assume that $\epsilon_a/k$ is at most of the order of unity, to keep a safe distance from any turning points of the potential. Then, to leading order in the WKB modes (\ref{wkbmode}) become
\be
f_{\pmb k, a}({\pmb x}) = \frac{D}{(2\pi)^{3/2} [1 - 4 (1 + \frac{m_0^2}{2k^2})\Phi]^{1/4}} e^{i {\pmb k} \cdot \int d {\pmb x} [\sqrt{1 - 4 (1 + \frac{m^2}{2k^2})\Phi} -\frac{2m_0 \epsilon_a}{k^2} \Phi]},  \label{wkbmode2}
\ee
i.e., they involve a small shift $\Delta \varphi$
\be
\Delta \varphi = - \frac{2 \epsilon_a}{v}\int d x \Phi \label{phaseshift}
\ee
 due to the different mass of the $a$-th degree of freedom; $x$ is the spatial coordinate in the direction of ${\pmb k}$ and $v = k/m_0$ is the velocity. Note the remarkable similarity of Eq. (\ref{phaseshift}) to the phase shift of massless particles.

 The relative phase shift $u$ for propagation along distance $L$ is
 \be
  u = \frac{\Delta \varphi }{kL} = - \frac{2 \epsilon_a m_0}{k^2} \langle \Phi\rangle,
 \ee
where $\langle \Phi\rangle = L^{-1}\int_0^L dx \Phi$ is the average of the potential along the line of propagation. Our analysis presupposes that $\frac{m_0}{k} \langle \Phi\rangle$ is at most of the order unity. As an example, we estimate the relative phase $u$ for a medium sized atom ($m_0 \sim 100$amu),  traveling a distance $10^4$km along the Earth-Moon axis.  We take $\epsilon_a = 1$eV, and we assume a velocity of $10^4m/s$. Then, $u \sim 10^{-12}$. The current relative accuracy achieved in atomic fountain clocks is of the order of $10^{-16}$  \cite{afc}. While it is far from obvious that the conditions of the experiment in Ref. \cite{afc} can be reproduced in a space environment, it appears that the detection of the phase shift (\ref{phaseshift}) is possible with current technologies.

The phase shift (\ref{phaseshift}) is a generalization of the phase shift of Ref. \cite{EPQS}, which was obtained for non-relativistic particles in a homogeneous gravitational field. The phase shift of \cite{EPQS} is at least an order of magnitude smaller than what can currently  be measured in Earth-based experiments. Our analysis makes it plausible that the generalization presented here---for relativistic particles in an inhomogeneous gravitational field---could be determined in space experiments. However, the exact prediction---rather than the order of magnitude estimate---depends on our model of particle detection and the particles' initial state.

The phase shift derived in Ref. \cite{EPQS} is intricately related to the equivalence principle for composite particles.
 We expect that  the phase shift (\ref{phaseshift}) derived here is also related to the equivalence principle for composite particles. However, the derivation presented here makes use of  WKB-type approximations, thus the results can only be  suggestive to the lowest order.  Any discussion of the equivalence principle of quantum systems must be made on the basis of exact solutions.

\section{Quantum EM field in  weak gravity}

\subsection{Electromagnetic field in curved spacetime}
The EM field in a spacetime $M$ with Lorentzian metric $g$ is expressed in terms of the two-form $F_{\mu \nu}$. Maxwell's equations are
\begin{eqnarray}
\partial_{\mu} \left( \sqrt{-g}F^{\mu \nu}\right) = 0, \hspace{0.6cm} \partial_{[\mu} F_{\nu \rho]} = 0.
\end{eqnarray}

Consider a static spacetime with metric (\ref{metric}).
We denote by $h$ the determinant of the three-metric, and by $E_{ijk} = \sqrt{h} \epsilon_{ijk}$ the covariant 3-d Levi-Civita symbol. We also define the electric field as $E^i = F^{0i}$ and the magnetic field as $B^i = \frac{1}{2}E^{ijk}F_{jk}$.
  Then, Maxwell's equations become
\begin{eqnarray}
\frac{\partial}{\partial t}\left(\sqrt{-g}E^i\right) - \partial _j(E^{ijk}  \sqrt{-g}  B_k) = 0, \label{Mxw1}\\
\frac{\partial}{\partial t}B^i + E^{ijk}  \partial_j (N^2 E_k) = 0, \label{Mxw2}\\
\partial_i(\sqrt{-g}E^i) = 0, \label{Mxw3}\\
\partial_i(\sqrt{h}B^i) = 0.
\end{eqnarray}
Eq. (\ref{Mxw1}) can also be written as
\be
N\dot{E}^i - E^{ijk}\partial_j (LB_k) = 0.
\ee

We  write an autonomous second order equation for the densitized electric field ${\cal E}^i = \sqrt{-g} E^i$,
\begin{eqnarray}
\ddot{\cal E}^i = \partial_j \left( L\sqrt{h} \nabla^j(\frac{L}{\sqrt{h}}{\cal E}^i) -  \nabla^i(\frac{L}{\sqrt{h}}{\cal E}^j)\right),
\label{KGG}
\end{eqnarray}
 The electric field  ${\cal E}^i$ also satisfies Gauss' law $\partial_i{\cal E}^i = 0$

To quantize, we look for   positive frequency solutions to Eq. (\ref{KGG}), i.e., solutions of the form $f^i_n (x) e^{- i \omega_n t}$, for $\omega_n > 0$ and some index $n$ that labels the solutions. They satisfy
\begin{eqnarray}
\partial_j \left( L\sqrt{h} \nabla^j(\frac{L}{\sqrt{h}}f_n^i) -  \nabla^i(\frac{L}{\sqrt{h}}f_n^j)\right) + \omega_n^2 f^i_n = 0, \label{WE} \\
\partial_if^i_n= 0 \label{WEb}
\end{eqnarray}

The field operator is $\hat{ E}^i(t, x) = \frac{1}{\sqrt{-g}} \sum_n \left[ \hat{a}^n f^i_n({\pmb r}) e^{- i \omega_n t} + \hat{a}^{\dagger}_n \bar{f}^i_n({\pmb r})e^{i \omega_n t}\right]$, in terms of the annihilation operators $\hat{a}^n$ and the creation operators $\hat{a}_n^{\dagger}$. The Wightman function for a general field state $|\Psi\rangle$ is
\begin{eqnarray}
G^{ij}_{(2)}(t, {\pmb x}; t', {\pmb x}') = 2 \mbox{Re} \left( \sum_{nm} \rho^n_b f^i_n({\pmb x}')\bar{f}^j_m({\pmb x})e^{-i (\omega_n t - \omega_m t')} \right)
\nonumber \\
 + \sum_a f^i_n({\pmb x}) \bar{f}^j_n({\pmb x}')e^{- i \omega_a(t-t')},
\end{eqnarray}
 where $\hat{\rho}^n_m = \langle \Psi|\hat{a}^{\dagger}_m \hat{a}^n|\Psi\rangle$ is the one-particle reduced density matrix associated to $|\Psi\rangle$.

 \subsection{Field modes}
 Next, we consider the case of a weak gravitational field that is a solution to Einstein's field equations in vacuum, Eq. (\ref{vacugr}).

To leading order in $\Phi$, Eq. (\ref{WE}) becomes
\begin{eqnarray}
\partial_j [ e^{4 \Phi}  \left(  \partial^j f^i - \partial^i  f^j +    3    f^i \partial^j \Phi -  3 f^j \partial^i \Phi \right) ] +\omega^2  f^i = 0. \label{WE2}\\
\partial_i f^i = 0. \label{WE2b}
\end{eqnarray}
where we dropped the index $a$ for brevity. All indices refer to the Cartesian coordinates, raised and lowered with the flat metric.

Eq.(\ref{WE2}) can be written as
\be
\nabla^2 f^i + J^{ijk}\partial_j f_k +q^2f^i = 0, \label{WE3}
\ee
 where
 \be
 J^{ijk} = 7 \delta^{ik} \partial^j\Phi - 3\delta^{jk} \partial^i \Phi - 4 \delta^{ij} \partial^k \Phi\\
 q^2 = 12 (\nabla \Phi)^2 + 3 \nabla^2 \Phi + \omega^2 e^{-4 \Phi}
 \ee

To solve Eq. (\ref{WE3}), we  substitute $f^i(x) = \Lambda^i_j(x) u^j(x)$, for some space-varying rotation matrix $ \Lambda$, which we choose so that the first order derivatives on $u$ cancel. This implies that
\be
\partial^k  \Lambda^i_j + \frac{1}{2} J^{ik}{}_l  \Lambda^l_j=0, \label{Sijr}
\ee
and that
\be
\nabla^2 u^l +( \Lambda^{-1})^l_k\nabla^2 \Lambda^k_j u^j +( \Lambda^{-1})^l_kJ^{kmn}\partial_m \Lambda_{nr}u^r + q^2 u^l = 0 \label{WE4}
\ee

In absence of gravity, $ \Lambda^{i}_j = \delta^{i}_j$, hence, for a weak gravitational field, we can write
\begin{eqnarray}
 \Lambda^i_j = \delta^i_j + \frac{1}{2} \int J^{i}{}_{kj} dx^k.  \label{Sijr}
\end{eqnarray}
In Eq. (\ref{WE4}), the term involving $J \partial  \Lambda$ is of order $\Phi^2$, hence, negligible. Again, to leading order in $\Phi$,
\be
( \Lambda^{-1})^l_k\nabla^2 \Lambda^k_j  \simeq \nabla^2 \Lambda^l_j = - 7 \partial^l\partial_j \Phi,
\ee
where we used the fact that $\nabla^2 \Phi = 0 $ in vacuum.

Hence,
\be
\nabla^2u^i + (q^2 \delta^i_j -7 \partial^i \partial_j \Phi)         u^j = 0,
\ee
where in this approximation $q^2 = \omega^2 e^{-4 \Phi}$. (The term $(\nabla \Phi)^2$ is of order $\Phi^2$.)

Let us denote the length-scale of second derivatives of $\Phi$ by $L$ ($L$ corresponds to some curvature radius of the spatial  Ricci scalar). Then if $\omega L >> 1$, the term $7 \partial^i \partial_j \Phi$ is much smaller than $q^2$ and it can be ignored. This condition is equivalent to the assertion that the wave-length $\lambda$ of radiation is much smaller than $L$. For the Newtonian potential of a mass $M$, $\Phi(r) =  -\frac{GM}{r}$, $L = r$. Since  in near-Earth experiments $r $ is larger than Earth's radius, the condition $\lambda << r$ is extremely well satisfied in the optical, microwave regime, up to radiowaves. Hence, we can write
\be
\nabla^2 u^i + q^2 u^i = 0. \label{KGEM}
\ee

Hence, each component of the mode function $u^i$ is proportional to $e^{\Phi + i {\pmb k} \cdot ( {\pmb x} + \delta {\pmb x})}$. In absence of gravity, such modes must be reduced to the usual modes of the electric field in Minkowski spacetime, namely, $-i \omega \epsilon^i e^{i {\pmb k} \cdot  {\pmb x}}$, where ${\pmb \epsilon} \cdot {\pmb k} = 0$.

Eq. (\ref{WE2b}) implies that $\partial_i \Lambda^{i}_j u^j + \Lambda^{i}_j \partial_iu^j = 0$. We expand $u^i = u^i_{(0)}+ u^i_{(1)}$, where $u^i_{(0)}$ is the mode function in absence of gravity and $u^i_{(1)}$, the correction of order $\Phi$. Then, we obtain
\be
\partial_iu^i_{(1)} = J^{i}{}_{ij}u^j_{(0)} - \frac{1}{2} \int dx^k \Theta^i_j \partial_iu^j_{(0)} \label{contr1}
\ee
where $\Theta^i_j = \int J^{i}_{kj}dx^k$.

The second term in the r.h.s. of Eq. (\ref{contr1}) is larger from the first by a factor of order  $\omega L$, where $L$ the typical distance traveled by light. Since $\omega L >> 1$ in space experiments, the first term in the r.h.s.of Eq. (\ref{contr1}) can be ignored.

Eq. (\ref{contr1}) describes a change in  the longitudinal part of $u^i$. It implements a correction of the form $\frac{k_i}{k} \beta e^{\Phi + i {\pmb k} \cdot ( {\pmb x} + \delta {\pmb x})}$, where $\beta$ is a slowly varying term, of the order of the ones that have been ignored in deriving Eq. (\ref{KGEM}). Then, Eq. (\ref{contr1}) implies that
\be
\beta = - \frac{k_i}{2k} \epsilon_j \Theta^i_j.
\ee
Hence,
\be
f^i _{{\pmb k}, \lambda}({\pmb x}) = \frac{D}{(2\pi)^{3/2}}(- i \omega_k) \left[ \epsilon^i + \frac{1}{2} \Theta^i_j \epsilon^j - \frac{1}{2}\Theta^k_j \frac{k^i k^j}{k^2} \epsilon_k\right]e^{\Phi + i {\pmb k} \cdot ( {\pmb x} + \delta {\pmb x})} \label{emmode}
\ee
where   the vector $\epsilon^i$ depends on the polarization $\lambda = 1, 2$.

Whenever we can ignore the polarization-changing term, the normalization constant $D = 1$, hence,
\be
f^i_{{\pmb k}, \lambda}({\pmb x}) =  \frac{1}{(2\pi)^{3/2}} (- i \omega_k) \epsilon_{\lambda}^i({\pmb k})   e^{\Phi + i {\pmb k} \cdot ( {\pmb x} - \delta {\pmb x})},
\ee
and {\em the gravitational field only acts as a dielectric that changes the optical path}, as explained in Sec. 2.4.

We also calculate
\be
|{\pmb f}_{{\pmb k}, \lambda}({\pmb x})|^2 = \frac{\omega_k^2}{(2\pi)^3} [1 + \theta_{{\pmb k},\lambda} ({\pmb x}) + 2 \Phi({\pmb x})],
\ee
where $\theta_{{\pmb k},\lambda}  := \Theta_{ij} \epsilon^i_{\lambda}({\pmb k})  \epsilon^i_{\lambda}({\pmb k}) $ is the size of the polarization rotation due to the gravitational field.

\section{Photodetection and particle detection}

Typical observables in interference experiments is the number of particles that are recorded by detectors. A detector follows a trajectory in spacetime, or a world-tube if we take its finite size into account. In Earth experiments, detectors are usually static with respect to the laboratory frame, while in deep space experiments detectors are expected to move along satellite orbits.

A QFTCST leads to predictions that can be tested in experiments only in conjunction with a particle detection theory, i.e., a modeling of the particle detectors that provides an expression for the detection probability for a detector in a specific trajectory as a function of time (or of the detector's proper time). An important  particle detection model is Glauber's photodetection theory \cite{Glauber}.  For a given quantum state $\hat{\rho}_0$ of the electromagnetic field, Glauber's theory expresses  the unnormalized  joint probability density $P_n(X_1, X_2, \ldots, X_n)$ for $n$ photodetection events at spacetime points $X_1, X_2, \ldots, X_n$ as
\begin{eqnarray}
P_n(X_1, X_2,\ldots, X_n) = Tr \left(\hat{E}^{i_n(+)}(X_n) \ldots \hat{E}^{i_2(+)}(X_2) \hat{E}^{i_1(+)}(X_1)  \hat{\rho}_0 \right.
\nonumber \\
\times \left.  \hat{E}^{(-)}_{i_1}(X_1)\hat{E}^{(-)}_{i_2}(X_2)\ldots \hat{E}^{(-)}_{i_n}(X_n) \right), \label{Glauber}
\end{eqnarray}
where $\hat{\pmb E}^{(\pm)}(X)$ is the positive (negative) frequency part of Heisenberg-picture operators that represent the electric field strength.

Glauber's theory has been immensely successful in quantum optics. However, it is limited in the following sense. First, it presupposes that all detectors are at rest in a given frame; a more detailed modeling of detectors is required if we want to identify the joint detection probability for detectors following different satellite orbits. Second, Glauber's theory involves a split of the field
into positive- and negative-frequency components. This split is non local and as such it could lead to non-causal behavior of the probabilities at large separations of the detectors. In this paper, we will use Glauber's theory for the EM field, because we will only be concerned with single detector measurements, i.e., with  $n = 1$ in Eq. (\ref{Glauber}). The construction of an  photodetection model, appropriate for deep space experiments, is undertaken in a forthcoming publication.

For a single detector, Eq. (\ref{Glauber}) yields
\be
P(X) =C  Tr\left( \hat{\rho}_0 \hat{\pmb E}^{(-)}(X) \cdot \hat{\pmb E}^{(+)}(X)\right), \label{Glauber1}
\ee
where $C$ is a normalization constant.

A more general particle detection model is provided by the Quantum Temporal Probabilities (QTP) method \cite{AnSav12, AnSav17, AnSav19, AnSav20}. The key idea in the QTP method is to distinguish between the time parameter of Schr\"odinger equation from the time variable associated to  particle detection \cite{Sav99, Sav10}. The latter time variable is then treated as a  macroscopic quasi-classical quantity  associated to the detector degrees of freedom. Hence, the method combines a microscopic modeling of the detector with a macroscopic description of its  measurement records,  expressed  in terms of classical spacetime coordinates.

For a massive scalar field with a single detection event, the QTP method leads to a probability formula \cite{AnSav19}
\be
P(X) = C \int d^4 Y K(Y) Tr \left[\hat{\rho}_0 \hat{\phi}(X+ \frac{1}{2}Y) \hat{\phi}(X - \frac{1}{2}Y)\right],
\ee
where $K(Y)$ is a symmetric kernel that contains all information about the detector. Hence, the detection probability is obtained from the knowledge of the field Wightman function.

For initial states that do not involve superpositions of particle number,
\be
P(X) = 2C \int d^4 Y K(Y) Tr \left[\hat{\rho}_0 \hat{\phi}^{(-)}(X+ \frac{1}{2}Y) \hat{\phi}^{(+)}(X - \frac{1}{2}Y)\right], \label{dettim}
\ee
where we split the field in its positive and negative frequency component. Then, an expression analogous to Eq. (\ref{Glauber1}) is obtained if we approximate $K(Y)$ by a delta function. In this paper, we only consider first order coherence, i.e., interference experiments that involve the reading of a single detector. In this context, Glauber's theory and QTP lead to the same results.

In general, QTP is not equivalent to Glauber's theory. Glauber's theory involves correlation functions that are defined with respect to a positive frequency split of the quantum fields. In contrast, probabilities in QTP are linear functions of correlation functions of the form
\be
G_n(X_1, X_2, \ldots, X_n; Y_1, Y_2, \ldots Y_n) = \langle \Psi|{\cal A} [\hat{\phi}(X_1) \hat{\phi}(X_2) \ldots \hat{\phi}(X_n)]\nonumber  \\ {\cal T} [\hat{\phi}(Y_n) \ldots \hat{\phi}(Y_2) \hat{\phi}(Y_1)]|\rangle \Psi\rangle,
\ee
where ${\cal T}$ stands for time order and ${\cal A}$ for anti-time order. Such correlation functions These correlation functions are not the ones for S-matrix theory but for real-time causal evolution. They involve both time-ordered and anti-time-ordered entries, as in the so-called Schwinger-Keldysh formalism \cite{SchwKeld}, now broadly used in many areas of physics from condensed matter physics to cosmology \cite{CaHu}. An analysis of the differences between Glauber theory and QTP in higher order correlations will be given in a different publication.

Here, we must point out that in space experiments, we are interested in real time properties of the fields and in regimes where the gravitational field is not switched-off at the detection point. Hence, the popular S-matrix formulation of QFT is not relevant here, the Schwinger-Keldysh formalism is preferable. This includes, for example, an analysis of non-unitary channels upon particle propagation, as in many models of gravitational decoherence

\section{Measurement schemes}
In this section, we present some measurement schemes relevant to quantum experiments in space. We focus on the case of photons, but analogous set-ups are in principle possible for matter waves. In all formulations, a particle detection theory is needed in order to obtain concrete predictions. Here, we mostly use Glauber's theory---except for the discussion of the time of arrival---keeping in mind that we mostly present simple set-ups where the difference from formalisms is not important. Such differences will be explored in a different publication.

  \subsection{Mach-Zehnder interferometry}

\begin{figure}
  \centering
 \includegraphics[width=0.65\textwidth]{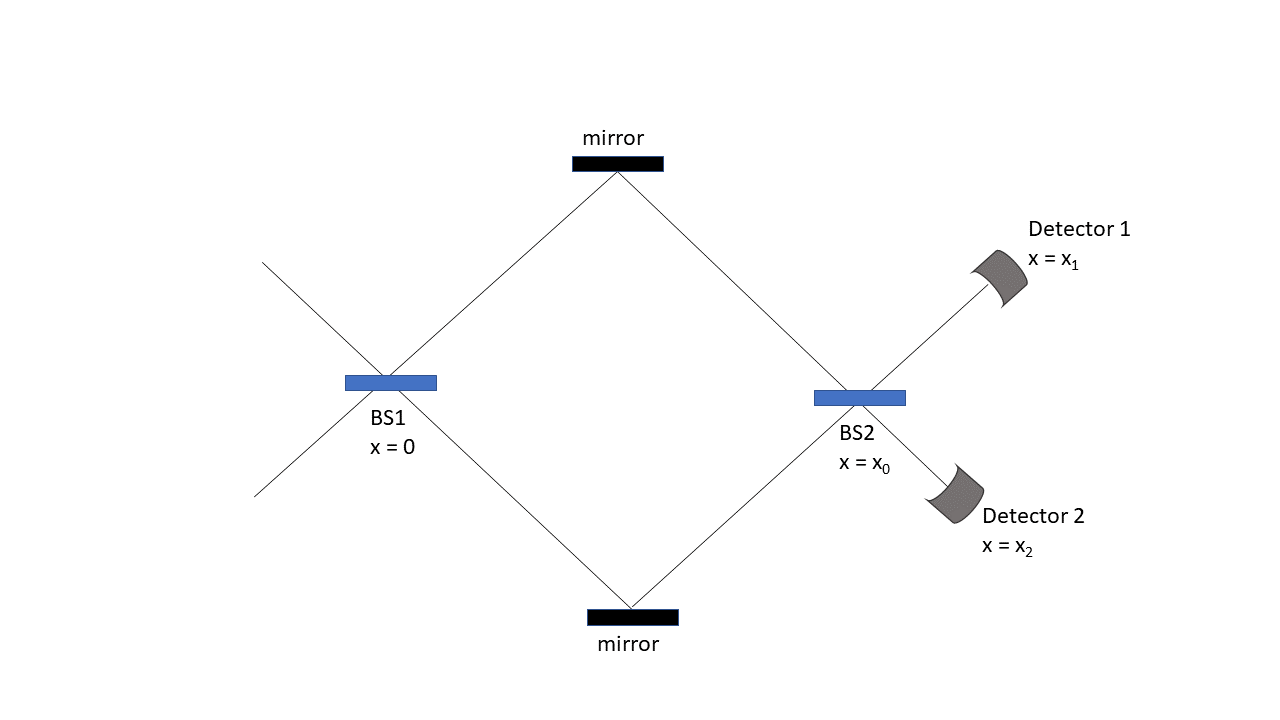}
    \caption{ An interferometer of the Mach-Zehnder type.}
\end{figure}

A Mach-Zehnder interferometer has two inputs and two outputs, each  corresponding to a different field mode. The two modes with ${\pmb k}_1$ and ${\pmb k}_2$   have the same frequency $\omega$ and the wave vectors are usually chosen normal to each other. Let us denote the associated mode function of type (\ref{emmode})  by ${\pmb f}_{1}$ and ${\pmb f}_{2}$, respectively.
The positive-frequency field operator then reads
\be
 \hat{\pmb E}^{(+)} ({\pmb x}, t) = [\hat{a}_1 {\pmb f}_1({\pmb x}) + \hat{a}_2 {\pmb f}_2({\pmb x})]e^{- i \omega t}.
\ee

A  beam splitter effects a transformation that corresponds to a matrix
\be
T = \frac{1}{\sqrt{2}} \left( \begin{array}{cc} 1&i\\i&1\end{array} \right).
\ee
We assume that the first beam splitter is located at ${\pmb x} = 0$, and choose a phase factor in  ${\pmb f}_i$, so that ${\pmb f}_1(0)= {\pmb f}_2(0)$. The beam splitter
 transforms  the field to
\be
 T \hat{\pmb E}^{(+)} ({\pmb x}, t) =  \frac{1}{\sqrt{2}}  \left[(\hat{a}_1 +  i \hat{a}_2 ) {\pmb f}_1({\pmb x}) + (i\hat{a}_1 + \hat{a}_2) {\pmb f}_2({\pmb x}) \right]e^{- i \omega t} ,
\ee
Let the second beam-splitter be located at ${\pmb x}_0$. The input is
\be
 \hat{\pmb E}^{(+)} ({\pmb x}_0, t) =  \frac{1}{\sqrt{2}}  \left[(\hat{a}_1 +  i \hat{a}_2 ) {\pmb f}_1({\pmb x}_0)  + (i\hat{a}_1 + \hat{a}_2) {\pmb f}_2({\pmb x}_0)) \right]e^{- i \omega t} ,
\ee
Let us write ${\pmb f}_2({\pmb x}_0) = {\pmb f}_1({\pmb x}_0)e^{i k \Delta L}$, where $\Delta L$ is the difference in optical length. The output from the second beam splitter is
\be
 T\hat{\pmb E}^{(+)} ({\pmb x}, t) =  \frac{1}{\sqrt{2}}  \left[(\hat{a}_1(1-e^{ik\Delta L})
 + i \hat{a}_2 (1+ e^{ik\Delta L})) {\pmb f}_1({\pmb x}) \right.
 \nonumber \\
 \left. +i \hat{a}_1 ((1+ e^{-ik\Delta L})  + \hat{a}_2(1-e^{-ik\Delta L}) ) {\pmb f}_2({\pmb x}) \right]e^{- i \omega t} .
\ee
Only the term proportional to $f_1$ contributes to photodetection at ${\pmb x}_1$, and similarly for  $ {\pmb x}_2$. Hence,

  \begin{eqnarray}
  \hat{\pmb E}^{(+)} ({\pmb x}_1, t)  = \frac{1}{\sqrt{2}}  \left[\hat{a}_1(1-e^{ik\Delta L})  +  i(1+ e^{ik\Delta L}) \hat{a}_2  \right] {\pmb f}_1({\pmb x}_1)e^{- i \omega t}  \\
    \hat{\pmb E}^{(+)} ({\pmb x}_2, t) = -\frac{1}{\sqrt{2}}  \left[i (1+ e^{ik\Delta L}) \hat{a}_1 + (1-e^{ik\Delta L}) \hat{a}_2 \right] {\pmb f}_2({\pmb x}_2)e^{- i \omega t}.
\ee
Hence, the detection probabilities are
\be
P_1:= P({\pmb x}_1, t) = 2C  |{\pmb f}_1({\pmb x}_1)|^2 \left[\sin^2 \frac{k\Delta L}{2} \langle\hat{a}_1^{\dagger} \hat{a}_1\rangle  + \cos^2 \frac{k\Delta L}{2} \langle\hat{a}_2^{\dagger}\hat{a}_2\rangle \nonumber \right. \\ \left. + \sin (k \Delta L) \mbox{Im} \langle\hat{a}_1^{\dagger}\hat{a}_2\rangle\right]\\
P_2:= P({\pmb x}_2, t) = 2C|{\pmb f}_2({\pmb x}_2)|^2 \left[\cos^2 \frac{k\Delta L}{2} \langle\hat{a}_1^{\dagger} \hat{a}_1\rangle    + \sin^2 \frac{k\Delta L}{2} \langle\hat{a}_2^{\dagger}\hat{a}_2\rangle \nonumber \right. \\ \left. - \sin (k \Delta L) \mbox{Im} \langle\hat{a}_1^{\dagger}\hat{a}_2\rangle\right]
\ee

\medskip

\noindent {\em Single incoming photon.}
 A common initial state in Mach-Zehnder interferometry is $|\Psi\rangle = |\psi\rangle\otimes |0\rangle$. Then,
\be
\delta P:= P_2 - P_1 = B_+ \langle\hat{a}_1^{\dagger} \hat{a}_1\rangle \cos (k \Delta L) + B_- \langle\hat{a}_1^{\dagger} \hat{a}_1\rangle,
\ee
where $B_{\pm} =  C(|{\pmb f}_1({\pmb x}_1)|^2 \pm |{\pmb f}_2({\pmb x}_2)|^2) $. Note that in flat space $B_-$ vanishes.

In the present system,
\be
\Delta L = \delta L -2 \oint \Phi[{\pmb x}(\lambda)]d\lambda, \label{deltaL}
\ee
where $\delta L$ is the difference in the `coordinate' length of the two paths (Minkowski distance), and the contour integral is over the loop formed by the two interferometric arms. The parameter $\lambda$ is the Minkowski length along the contour.

A key point here is that $\delta L$ is the difference in length between the two paths, as measured with respect to the flat Minkowski metric. If we consider the difference $\delta S$ with respect to the full metric, then $\delta S = \delta L - \oint d\lambda \Phi[{\pmb x}(\lambda)]$, hence,
\be
\Delta L = \delta S - \oint \Phi[{\pmb x}(\lambda)]d\lambda.
\ee
Unlike laboratory experiments, it is not obvious that we can fix $\delta S$ to be zero. The reason is that we cannot switch off the gravitational field, and any photons employed to fix the distances between the elements of the interferometer will `see' the optical metric, and hence, the path difference $\Delta L$. To determine $\delta L$ or $\delta S$, we would have to use methods that do not involve photons, for example, time of flight measurements of neutrons or electrons.

Consider a homogeneous field $\Phi= -g x$ and a rectangular  interferometric path of length $h$ in the $x$ direction and of length $d$ in the $y$ direction.
Then,
\begin{eqnarray}
\Delta L = \delta L + 2  g h d,
\end{eqnarray}
in agreement with Ref. \cite{Terno}.

Note that the contrast $V$ of interference in $\delta P$ defined as $V = \frac{\delta p_{max} - \delta p_{min}}{ \delta p_{max} + \delta p_{min}}$ is
\be
V = \frac{B_+ - B_-}{B_+ + B_-} = \frac{|{\pmb f}_1({\pmb x}_1)|^2}{|{\pmb f}_2({\pmb x}_2)|^2} = 1  +  \theta_{1} ({\pmb x}_1) -  \theta_{2} ({\pmb x}_2)   + 2 \Phi({\pmb x}_1) - 2 \Phi({\pmb x}_2)
\ee
The contrast differs appreciably from unity  (i) if the detectors are located at points with very different value of the potential, or  (ii) if the rotation of the polarization along the two paths is significant.

\subsection{Hong-Ou-Mandel interferometer}

\begin{figure}
    \centering
\includegraphics[width=0.65\textwidth]{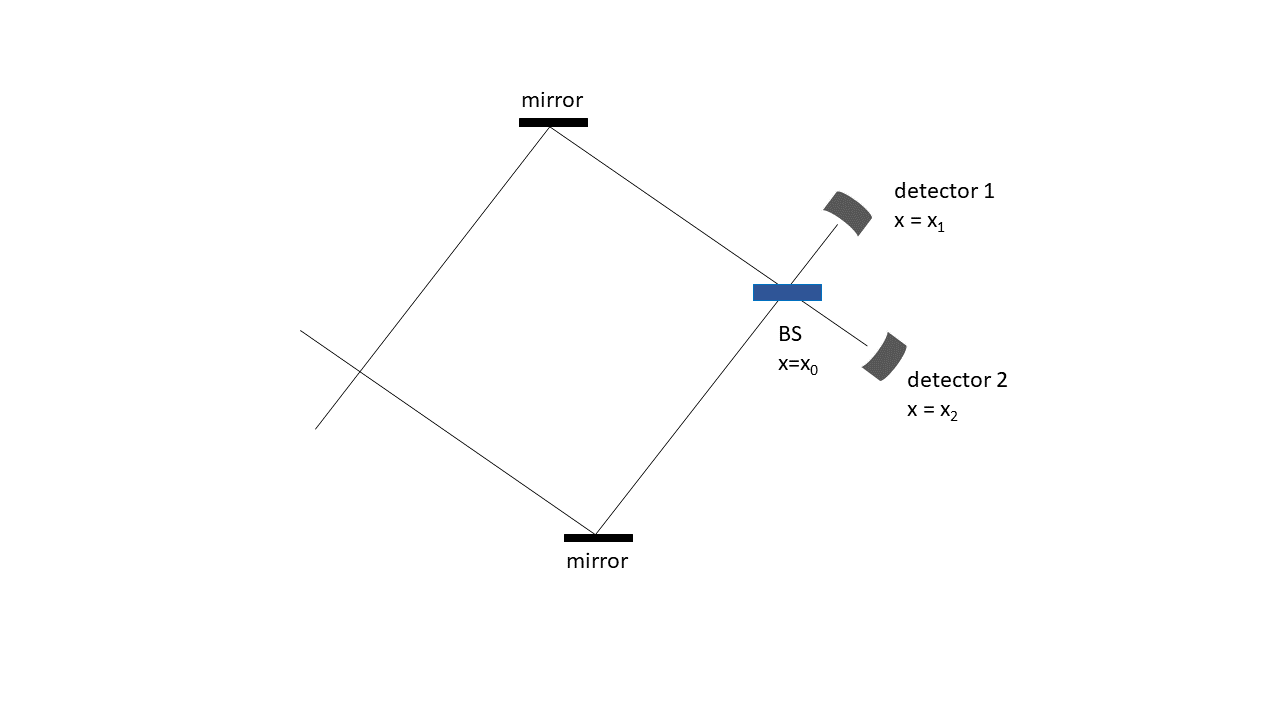}
    \caption{ An interfereometer of the Hong-Ou-Mandel type.}
 \end{figure}

In a Hong-Ou-Mandel interferometer, the input consists again of two modes $1$ and $2$, leading to a  positive-frequency field operator
\be
 \hat{\pmb E}^{(+)} ({\pmb x}, t) = [\hat{a}_1 {\pmb f}_1({\pmb x}) + \hat{a}_2 {\pmb f}_2({\pmb x})]e^{- i \omega t}.
\ee
There is a single beam splitter at ${\pmb x} = {\pmb x}_0$. Again, we assume  that ${\pmb f}_2({\pmb x}_0) = {\pmb f}_1({\pmb x}_0)e^{i k \Delta L}$, where $\Delta L$ is the difference in optical length. Then, the output from the beam splitter is
\be
 T\hat{\pmb E}^{(+)} ({\pmb x}, t) =  \frac{1}{\sqrt{2}}  \left[(\hat{a}_1
 + i \hat{a}_2 e^{ik\Delta L}) {\pmb f}_1({\pmb x})
 +(i \hat{a}_1  e^{-ik\Delta L}  + \hat{a}_2 ) {\pmb f}_2({\pmb x}) \right]e^{- i \omega t} . \hspace{2cm}
\ee

The detection probability at detectors 1 and 2 are
\be
P_1 = C|{\pmb f}_1({\pmb x}_1)|^2  \left[  \langle\hat{a}_1^{\dagger} \hat{a}_1\rangle +  \langle\hat{a}_2^{\dagger} \hat{a}_2\rangle + i  \langle\hat{a}_1^{\dagger} \hat{a}_2\rangle e^{ik \Delta L} - i  \langle\hat{a}_2^{\dagger} \hat{a}_1\rangle^{- i k \Delta L} \right] \\
P_2 = C|{\pmb f}_2({\pmb x}_2)|^2  \left[  \langle\hat{a}_1^{\dagger} \hat{a}_1\rangle +  \langle\hat{a}_2^{\dagger} \hat{a}_2\rangle - i  \langle\hat{a}_1^{\dagger} \hat{a}_2\rangle e^{ik \Delta L} +i  \langle\hat{a}_2^{\dagger} \hat{a}_1\rangle^{- i k \Delta L} \right]
\ee

Consider a general pure two-mode state
\be
|\Psi\rangle = \frac{1}{2} (1+c^2)^{-1/2} [(\hat{a}^{\dagger}_1)^2 +  (\hat{a}^{\dagger}_2)^2 + 2c e^{i \chi} \hat{a}^{\dagger}_1 \hat{a}^{\dagger}_2]|0\rangle,
\ee
with $c > 0$, and $\chi$ a phase. For this state, $\langle\hat{a}_1^{\dagger} \hat{a}_1\rangle = \langle\hat{a}_2^{\dagger} \hat{a}_2\rangle = 1$ and $\langle\hat{a}_1^{\dagger} \hat{a}_2\rangle = \frac{2c}{1+c^2} \cos \chi$,
so that
\be
P_1 = 2C|{\pmb f}_1({\pmb x}_1)|^2  \left(  1  - \frac{2c}{1+c^2} \cos \chi \cos (k \Delta L)\right) \\
P_2 = 2C|{\pmb f}_2({\pmb x}_2)|^2 \left(  1 +  \frac{2c}{1+c^2} \cos \chi \cos (k \Delta L)\right)
\ee

We note that for $c = 1$ and $\chi = 0$, $P_1$ vanishes for $\Delta L = 0$.  In this case, the presence of a signal on the first detector is a sign of non-zero $\Delta L$.

\subsection{Time of arrival measurements}
In a time-of-arrival measurement, the observable is the time $t$ it takes for a particle  to arrive at a detector located at distance $L$ from the source. For sufficiently large $L$, only particles with momentum along the line connecting the source to the detector. Hence, without loss of generality, we can consider an one dimensional problem, i.e., restrict to coordinates $t$ and $x$.

Let the detector be located at $x = L$ in the Minkowski coordinates. We will follow the analysis of Ref. \cite{AnSav19} for the time-of-arrival of scalar particles, given by
 Eq. (\ref{dettim}) with an arbitrary function $K(t, x)$ describing the detector's response. We will take $m = 0$, as to provide an simple model of photon time-of-arrival measurements. Our aim is not to calculate an explicit time-of-arrival probability, but to see how the phase shift is manifested in time of arrival measurements.

  We find,
\be
P(L, t) = \int dk dk' \rho_0(k, k')
\nonumber \\
\int dt dy K(t, y)[f_k(L+ \frac{1}{2}y, t+\frac{1}{2}s ) f^*_{k'}(L+ \frac{1}{2}y, t+\frac{1}{2}s )] \label{plt}
\ee
In one spatial dimension, and for $m = 0$, the mode functions (\ref{modesol}) become
\be
f_k(x) = \frac{1}{\sqrt{2\pi}} e^{  i  k \cdot ( x - \delta  x)}.
\ee

Suppose that the size of the detector is much smaller than the typical length scale $L$ of variation in $\Phi$. Then, we can approximate $\delta x(L \pm \frac{1}{2}y) \simeq \delta x(L)$ and  Eq. (\ref{plt}) yields
\be
P(L, t) = P_0[L - \delta x(L)], \label{timeofarriv}
\ee
where $P_0$ is the time-of-arrival probability density for free particles. Hence, $\delta x(L)$ is the change in the time of arrival: the gravitational field actually slows the effective speed of light just like a dielectric. This is an important point, because in General Relativity distances are in principle determined by light rays and clocks that measure emission and arrival times for local observers \cite{EPS73}. Eq. (\ref{timeofarriv}) implies that if we use light rays in order to determine the distance between the elements of the interferometer, we will always come up with the length difference $\Delta L$ of Eq. (\ref{deltaL}). It is impossible to identify the gravitational phase difference using only photons.

In this paper, we consider only the case of $m = 0$, where Eq. (\ref{timeofarriv}) can be derived without any specific modeling assumptions. The case $m \neq 0$ is more complex. The spread of the distribution is sensitive on the modeling of the apparatus, and the presence of gravity introduces additional complications. We shall come to this in a future publication.

\subsection{Measurement of the phase shift from internal degrees of freedom}
Here, we will undertake a simplified description of detection of particles with internal degrees of freedom. As in the interferometry experiments of Sec. 5.1, we will assume that only a finite number of modes is excited and we will ignore wave packets. We assume that all modes have momenta ${\pmb k}$ and that they differ only in the internal degrees of freedom.

We write $\hat{\phi}^{(+)}(t, {\pmb x}) = \sum_a   \hat{a}_{{\pmb k}, a} f_{{\pmb k}, a}({\pmb x}) e^{- i \omega_{{\pmb k}, a} t}$, where $\omega_a = \sqrt{{\pmb k}^2 + m_a^2} \simeq \omega_0 + \frac{\epsilon_a m_0 }{\omega_0}$. We use an analogue of Glauber's formula for detection probabilities
\be
P(t, {\pmb x}) = \sum_{ab} \rho_{ab}({\pmb k}) |f_{\pmb k}({\pmb x})|^2 e^{i \frac{m_0(\epsilon_a - \epsilon_b)}{k^2} {\pmb k} \cdot {\delta {\pmb Y}}  }. \label{detprobi}
\ee
Here,  $\rho_{ab} ({\pmb k}) = \langle \hat{a}_b^{\dagger}({\pmb k}) \hat{a}_a({\pmb k})\rangle$ is the reduced density matrix for the internal degrees of freedom, and
\be
\delta {\pmb Y} =  \int d {\pmb x} \Phi - {\pmb v} t,
\ee
where ${\pmb v} = {\pmb k}/m_0$ is the velocity vector.

Eq. (\ref{detprobi}) strongly suggests that the phase shift (\ref{phaseshift}) is measurable. The only requirement is that the reduced density matrix $\rho_{ab} ({\pmb k})$ is not diagonal, i.e., that the system has been prepared in a superposition of the eigenstates of the Hamiltonian for the internal degrees of freedom.

A precise quantitative prediction requires a more elaborate analysis. There are three reasons. First, the elements $\rho_{ab}$ may very well include an oscillatory phase $e^{i s({\pmb k}) (\epsilon_a- \epsilon_b)}$ for some constant $s$, thus changing the oscillating phase by a ${\pmb k}$-dependent factor $s$. Such a phase factor may arise from an external interaction that allows us to prepare the system in a non-diagonal $\rho_{ab} ({\pmb k})$. Hence, the preparation of the system must be specified.

Second, the oscillating phase contains a time factor $t$; the probabilities with respect to the time $t$ are peaked at the time where the wave-packet reaches the detector, but this time depends on ${\pmb k}$, and thus on the momentum distribution of the wave-packet.
Third, as shown in Ref.  \cite{AnSav12}, decoherence effects on the detector may lead to the suppression of some (but not all) interference terms, and thus, changing the effective phase shift.

The fact that the gravity induced phase shift (\ref{phaseshift})  does not depend on the particle's mass, but only on its velocity,  is a manifestation of the equivalence principle for quantum systems, as explained in Ref. \cite{EPQS}. Our analysis here highlights the necessity of a precise QFTCST treatment of particles for understanding such phenomena, but also the fact that a proper model for particle detection is indispensable to making numerically accurate physical predictions.

\section{Conclusion}

Currently planned deep-space experiments will test regimes where QFTCST is relevant. The importance of QFTCST is manifested at three levels. First, QFTCST is the only consistent theory that we have describing the evolution of quantum systems in the presence of a background gravitational field. Hence, it is indispensable for the theoretical description of precision measurements, for understanding the equivalence principle for quantum systems, and as a final arbitrator of any prediction obtained through other methods.

Second, physical predictions for measurements at large separations or ones involving fast moving detectors  require a photo-detection theory or a particle detection theory. Such theories can only be formulated in terms of quantum fields, and they must be tested with respect to the properties of causality and locality.  Third, in the medium term, space experiments will allow us to explore long-standing issues in the foundations of QFT, but also to formulate a genuinely relativistic theory of quantum information based on quantum fields.

This paper touches upon the first two of these key points. We use the minimal tools from QFTCST that are necessary in order to describe planned experiments, as our primary aim is to build towards a formulation of a relativistic quantum optics theory in the presence of gravity. We also analyzed the behavior of composite massive particles in an inhomogenous gravitational field, and we identified a gravitational phase shift due to the internal degrees of freedom. Finally, we showed how a consistent  detection theory (both for photons and massive relativistic particles) is essential for meaningful physical predictions. Experiments involving distant correlations, to be described in later publications, will allow us to distinguish between competing detection theories. \\

\noindent{\bf Acknowledgments.}  We have benefited from a long term discussion of those topics with Albert Roura and also from discussions in the context of the DSQL collaboration, especially questions and/or remarks from Makan Mohageg, Paul Kwiat and Paolo Villoresi.
 This research was partially supported by a grant from the Julian Schwinger Foundation. BLH is supported by NASA/JPL grant 301699-00001.


\section*{References}
\begin{thebibliography}{}

\bibitem{EPQS} C. Anastopoulos and Bei-Lok Hu, {\em Equivalence Principle for Quantum Systems: Dephasing and Phase Shift of Free-Falling Particles}, Class. Quant. Gravity  35,  035011 (2018).

\bibitem{DSQL} DSQL https://techport.nasa.gov/view/94990

\bibitem{Makan}  M. Mohageg, C. Anastopoulos, J. Gallicchio,  B. L. Hu, T. Jennewein, P. Kwiat,  S-Y Lin et al (22 authors), {\em The Deep Space Quantum Link: Prospective Fundamental Physics Experiments using Long Baseline Quantum Optics} (in preparation).

\bibitem{Terno} D. R. Terno, G. Vallone, F. Vedovato, and P. Villoresi, {\em Large-Scale Optical Interferometry in General Spacetimes},  Phys. Rev. D101, 104052 (2020).

D. R. Terno, F. Vedovato, M. Schiavon, A. R. Smith, P. Magnani, G. Vallone,  and P. Villoresi, {\em  Proposal for an Optical tTest of the Einstein Equivalence Principle}, arXiv:1811.04835 (2018).

\bibitem{Rideout} D. Rideout, T. Jennewein, G. Amelino-Camelia, T. F. Demarie, B. L. Higgins, A. Kempf,  A. Kent, R. Laflamme, X.  Ma, R. B. Mann, and E. Martin-Martinez, {\em Fundamental Quantum Optics Experiments Conceivable with Satellites—Reaching Relativistic Distances and Velocities}, Class. Quant. Grav.  29, 224011 (2012).

\bibitem{Ohlsson} T. Ohlsson,  {\em Relativistic Quantum Physics: From Advanced Quantum Mechanics to Introductory Quantum Field Theory} (Cambridge University Press, 2012).

\bibitem{BirDav} N. D. Birrell and P. Davies, {\em Quantum Fields in Curved Space} (Cambridge University Press, 1982).

\bibitem{Wald} R. M. Wald, {\em Quantum Field Theory in Curved Spacetime and Black Hole Thermodynamics } (University of Chicago Press, 1994).

\bibitem{Fulling} S. A. Fulling, {\em Aspects of Quantum Field Theory in Curved Spacetime} (Cambridge University Press, 1989).

\bibitem{Will}  C.M. Will, {\em  The Confrontation between General Relativity and Experiment}, Liv. Rev.  Relat. 17, 1 (2014).

\bibitem{COW}  R. Colella, A. W. Overhauser, and S. A. Werner, {\em Observation of Gravitationally Induced Quantum Interference}, Phys. Rev. Lett. 34, 1472 (1975).


\bibitem{Borde} Ch. J. Bord\'e,  {Theoretical Tools for Atom Optics and Interferometry}, C. R. Acad. Sci. Paris, t. 2, Ser. IV, 509 (2001).

C. L\"ammerzahl and Ch. J. Bord\'e, {\em Atom Interferometry in Gravitational Fields: Influence of Gravitation on the Beam Splitter}, Gen. Rel. Grav. 31, 635 (1999).

\bibitem{LamBor}C. L\"ammerzahl and Ch. J. Bord\'e, {\em Rabi Oscillations in Gravitational Fields: Exact Solution}, Phys. Lett. A203, 59 (1995).

\bibitem{MarAud}  K.-P. Marzlin and J. Audretsch, {\em ‘‘Freely’’ Falling Two-Level Atom in a Running Laser Wave}, Phys. Rev. A53, 1004 (1993).

\bibitem{Roura} A. Roura, {\em    Gravitational Redshift in Quantum-Clock Interferometry}, Phys. Rev. X10, 021014 (2020).

\bibitem{HehlNi}  F. W. Hehl, W.-T.  Ni, {\em Inertial Effects of a Dirac Particle}, Phys. Rev. D42, 2045 (1990).

\bibitem{neutron} H. Rauch  and S. A. Werner, {\em  Neutron Interferometry: Lessons in Experimental Quantum Mechanics, Wave-Particle Duality, and Entanglement (Vol. 12)}  (Oxford University Press, 2015).

\bibitem{Ana} J. Anandan,  {\em Curvature Effects in Interferometry}, Phys.  Rev. D30, 1615 (1984).

\bibitem{Abele} H. Abele, {\em The Neutron. Its Properties and Basic Interactions}, Prog. Part.   Nucl. Phys. 60, 1 (2008).

H. Abele and H. Leeb, {\em Gravitation and Quantum Interference Experiments with Neutrons}, New J. Phys. 14, 055010 (2012).


\bibitem{Ryder} K. Varju and L. H. Ryder, {\em General Relativistic Treatment
of the Colella- Overhauser- Werner Experiment on Neutron
Interference in a Gravitational Field}, Am. J. Phys. 68,
404409 (2000)

A. Galiautdinov and L. H. Ryder, {\em  Neutron Interference in the Earth’s Gravitational Field},  Gen. Rel. Grav. 49, 82 (2017).

\bibitem{Green} D. M. Greenberger, W. P. Schleich,  and E. M. Rasel, {\em Relativistic Effects in Atom and Neutron Interferometry and the Differences between Them}, Phys. Rev. A86, 063622 (2012).

\bibitem{Mash} B. Mashhoon, {\em  Neutron Interferometry in a Rotating Frame of Reference}, Phys. Rev. Lett.  61, 2639 (1988).

 C. Chicone, B. Mashhoon,  and B. Punsly, {\em  Relativistic Motion of Spinning Particles in a Gravitational Field},  Phys. Lett. A343, 1 (2005).

\bibitem{KhrPom} I. B. Khriplovich and A. A. Pomeransky, {\em Equations of Motion of Spinning Relativistic Particle in external Fields}, J. Exp. Theor. Phys. 86, 839 (1998).

\bibitem{Alsing} P. M. Alsing, I. Fuentes-Schuller, R. B. Mann, and T. E. Tessier, {\em  Entanglement of Dirac Fields in Noninertial Frames}, Phys. Rev. A, 74, 032326 (2006).

\bibitem{RelQOp1}  E. Martín-Martínez and P. Rodriguez-Lopez, {\em Relativistic Quantum Optics: The Relativistic Invariance of the Light-Matter Interaction Models},  Phys. Rev. D97, 105026 (2018).

\bibitem{RelQOp2}T. Kolioni and C. Anastopoulos, {\em Detectors Interacting through Quantum Fields: Non-Markovian Effects, Nonperturbative Generation of Correlations, and Apparent Noncausality},  Phys. Rev. A 102, 062207 (2020).

\bibitem{RelQOp3} C. J. Fewster  and R. Verch, {\em Quantum Fields and Local Measurements}, Comm. Math. Phys. 378, 851 (2020).

\bibitem{Glauber}  R. J. Glauber, {\em The Quantum Theory of Optical Coherence}, Phys. Rev. 130, 2529 (1963);  {\em Coherent and Incoherent States of the Radiation Field}, Phys. Rev. 131, 2766 (1963).

 \bibitem{RWA} Chris Fleming, Nicholas Cummings,  C. Anastopoulos  and B. L. Hu,  {\em The Rotating Wave Approximation: Consistency and Applicability from a Quantum Open Systems Analysis},     J. Phys. A: Math. Theor. 43  (2010) 405304.

\bibitem{Schl71} S. Schlieder,  {\em Zum Kausalen Verhalten eines Relativistischen Quantenmechanischen Systems}, in
"Quanten und Felder, W. Heisenberg zum 70. Geburtstag",  Ed. H.P. D\"urr, (Vieweg 1971) p. 145.

\bibitem{Heg98} G. C.  Hegerfeldt,  {\em Instantaneous Spreading and Einstein Causality in Quantum Theory},
Annalen der Physik 7, 716 (1998).


\bibitem{Mal96} D. Malament,  {\em In Defense of Dogma: Why There Cannot
Be A Relativistic Quantum Mechanics of (Localizable) Particles}, in
"Perspectives on Quantum Reality",
ed. R. Clifton  (Kluwer Academic, Dordrecht 1996).


\bibitem{Fermi} E. Fermi, {\em Quantum Theory of Radiation}, Rev. Mod. Phys. 4, 87 (1932).

\bibitem{Shiro} M. I. Shirokov, {\em Velocity of Electromagnetic Radiation in Quantum Electrodynamics}, Yad.Fiz. 4, 1077 (1966)[Sov. J. Nucl. Phys. 4 ,774(1967)].

\bibitem{Fermiproblem} B. Ferretti, {\em Propagation of Signals and Particles} "Old and New Problems in Elementary Particles" ed. by G. Puppi, (Academic Press, New York 1968).

  P.W. Milonni and P.L. Knight, {\em Retardation in the Resonant Interaction of Two Identical Atoms}, Phys. Rev. Lett. A10, 1096 (1974).

   M. H. Rubin, {\em Violation of Einstein Causality in a Model Quantum System}, Phys. Rev. D35, 3836 (1987).

    A.K. Biswas, G. Compagno, G.M. Palma, R. Passante, and R. Persico, {\em Virtual Photons and Causality in the Dynamics of a Pair of Two-Level Atoms}, Phys. Rev. Lett. A42, 4291 (1990).

     A.  Valentini, {\em Non-local Correlations in Quantum Electrodynamics}, Phys. Rev. Lett. A153, 321 (1991).

\bibitem{Heg} G.C. Hegerfeldt,{\em Causality Problems for Fermi's Two-Atom System},  Phys. Rev. Lett. 72, 596 (1994).

\bibitem{Heg2} G. C. Hegerfeldt, {\em Problems about Causality in Fermi'sTwo-Atom Model and Possible Resolutions}, in "Non-Linear, Deformed and Irreversible Quantum Systems", eds: H.-D. Doebner, V. K. Dobrev, P. Nattermann (World Scientific, Singapore 1995).

\bibitem{Buch} D Buchholz and J Yngvason, {\em There are no Causality Problems for Fermi's Two-Atom System}, Phys. Rev. Lett. 73, 613 (1994).

\bibitem{causalQI} 	O. Oreshkov et al, {\em Quantum Correlations with no Causal order}, Nature Communications 3, 1092 (2012).

     G Chiribella et al, {\em Quantum Computations without Definite Causal Structure}, Phys. Rev. A88, 022318 (2013).

     \bibitem{Relcaus} S. Popescu and L. Vaidman,  {\em Causality Constraints on Nonlocal Quantum Measurements},
Phys. Rev. A 49, 4331 (1994).

 A. Peres,  {\em Classical Interventions in Quantum Systems. II. Relativistic Invariance}, Phys. Rev. A 61, 022117 (2000).

D. Beckman, D. Gottesman, M. A. Nielsen, and J. Preskill,  {\em Causal and Localizable Quantum Operations}, Phys. Rev. A64,  052309 (2001).

\bibitem{OnVi} L. Viola and R. Onofrio, {\em Testing the Equivalence Principle Through Freely Falling Quantum Objects}, Phys. Rev. D55, 455 (1997).
    
    \bibitem{ZyBr} M. Zych and C. Bruckner, {\em Quantum Formulation of the Einstein Equivalence Principle}, Nature Physics 14,  1027 (2018).


\bibitem{AnHu15} C. Anastopoulos and B. L. Hu, {\em Probing a Gravitational Cat State}, Class. Quant. Grav. 32, 165022 (2015).

\bibitem{Bose17} S. Bose, A. Mazumdar, G. W. Morley, H. Ulbricht, M. Toro, M. Paternostro, A. A. Geraci, P. F. Barker, M. S. Kim, and G. Milburn, {\em A Spin Entanglement Witness for Quantum Gravity}, Phys. Rev. Lett. 119, 240401 (2017).

\bibitem{Vedral17} C. Marletto and V. Vedral, {\em Gravitationally Induced Entanglement between Two Massive Particles is Sufficient Evidence of Quantum Effects in Gravity}, Phys. Rev. Lett. 119, 240402 (2017).

\bibitem{AnHucr} C. Anastopoulos and B. L. Hu, {\em Comment on "A Spin Entanglement Witness for Quantum Gravity" and on "Gravitationally Induced Entanglement between Two Massive Particles is Sufficient Evidence of Quantum Effects in Gravity"}, arXiv:1804.11315.  For an extended version containing critical exchanges, see   https://www.researchgate.net/publication/324859896.

    \bibitem{AnHu20} C. Anastopoulos and B. L. Hu, {\em Quantum Superposition of Two Gravitational Cat States}, Class. Quant. Grav. 37, 235012 (2020).


\bibitem{ZCPB} M. Zych, F. Costa, I. Pikovski, and C. Brukner,  {\em Bell’s Theorem for Temporal Order}, Nat. Commun. 10, 3772 (2019).


\bibitem{ALS21} C. Anastopoulos, M. Lagouvardos and K. Savvidou, {\em Gravitational Effects in Macroscopic Quantum Systems: a First-Principles Analysis}, Class. Quant. Grav. (2021), in press. https://doi.org/10.1088/1361-6382/ac0bf9



\bibitem{AnalogG} See, for example, C. Barcelo, S. Liberati,  and M. Visser, {\em  Analogue Gravity}, Liv. Rev.  Relat. 14, 1 (2011).

\bibitem{GQP} See, for example, papers collected in:  M. Aspelmeyer, C. Brukner, D. Giulini,  and G. Milburn, {\em  Focus on Gravitational Quantum Physics} New J. Phys., 19, 050401 (2017).

\bibitem{RQI} See, for example, the International Society on Relativistic Quantum Information, \url{www.isrqi.net}.


\bibitem{Skrotski} G. V. Skrotskii, {\em The Influence of Gravitation on the Propagation of Light}, Sov. Phys. Doklady 2, 226  (1957); A. M. Volkov, A. A. Izmest’ev, and G. V. Skrotskii, {\em The Propagation of Electromagnetic Waves in a Riemannian Space}, Sov. Phys. JETP 32, 686 (1971).

\bibitem{Plebanski}J. Plebanski, {\em
Electromagnetic Waves in Gravitational Fields }, Phys. Rev. 118, 1396 (1960).

\bibitem{defelice} F. de Felice, {\em On the Gravitational Field Acting as an Optical Medium}, Gen. Rel. Grav. 2, 347 (1971).

\bibitem{terno2} A. Brodutch, A. Gilchrist, T. Guff, A. R. H. Smith, and D. R. Terno, {\em Post-Newtonian Gravitational Effects in Optical Interferometry}, Phys. Rev. D 91, 064041 (2015).



\bibitem{AnSav12} C. Anastopoulos and N. Savvidou,  {\em Time-of-Arrival Probabilities for General Particle Detectors}, Phys. Rev. A86, 012111 (2012).

\bibitem{AnSav19}  C. Anastopoulos and N. Savvidou, {\em Time of Arrival and Localization of Relativistic Particles}, J. Math. Phys. 60, 0323301 (2019).


\bibitem{ACL88} M. A. Abramowicz, B. Carter, J. P. Lasota, {\em  Optical Reference Geometry for Stationary and Static Dynamics}, Gen. Rel. Grav. 20, 1173 (1988).

\bibitem{polar}  F. Fayos and J. Llosa, {\em Gravitational Effects on the Polarization Plane}, Gen. Rel. Grav.14, 865 (1982);
  A. Brodutch and D. R. Terno, {\em Polarization Rotation, Reference Frames,  and  Machs  Principle},  Phys.  Rev.  D84,  121501(R)(2011).

\bibitem{AnSav17} C. Anastopoulos and N. Savvidou, {\em Time-of-Arrival Correlations}, Phys. Rev. A95, 032105 (2017).

\bibitem{AnSav20}  C. Anastopoulos and N. Savvidou, {\em  Multi-Time Measurements in Hawking Radiation: Information at Higher-Order Correlations }, Class. Quantum Grav. 37, 025015 (2020).



    \bibitem{Sav99} K. Savvidou,  {\em The Action Operator for Continuous-time Histories}  J. Math. Phys. 40, 5657 (1999); {\em Continuous Time in Consistent Histories}, gr-qc/9912076.

\bibitem{Sav10} N. Savvidou, {\em Space-time Symmetries in  Histories Canonical Gravity}, in "Approaches to Quantum Gravity", edited by D. Oriti (Cambridge University Press, Cambridge 2009).

\bibitem{afc} T. P. Heavner, E. A. Donley, F. Levi, G. Costanzo, T. E. Parker, J. H. Shirley, N. Ashby, S.Barlow and S. R. Jefferts, {\em First Accuracy Evaluation of NIST-F2}, Metrologia 51, 174 (2014).

\bibitem{SchwKeld}  J. S. Schwinger, {\em Brownian Motion of a Quantum Oscillator}, J. Math. Phys. 2, 407 (1961).

 L. V. Keldysh, {\em
Diagram Technique for Nonequilibrium Processes}, Zh. Eksp. Teor. Fiz. 47, 1515 (1964).

\bibitem{CaHu} E. Calzetta and B. L. Hu, {\em Nonequilibrium Quantum Field Theory} (Cambridge University Press, 2008).
    
\bibitem{EPS73} J. Ehlers, F. A. E. Pirani,  and A. Schild, {\em The Geometry of Free Fall and Light Propagation}, p. 63 in O'Raifeartaigh, L. (ed.), "General Relativity, Papers in Honor of J. L. Synge"   (Clarendon Press, Oxford 1972).
\end{thebibliography}
\end{document}